\documentstyle[prl,aps,preprint,epsf]{revtex}
\tightenlines

\begin{document}
\draft

\title{
A Theoretical Study on Spin-Dependent Transport of 
``Ferromagnet/Carbon Nanotube Encapsulating Magnetic Atoms/Ferromagnet'' 
Junctions with 4-Valued Conductances
}
\author{
Satoshi~\textsc{Kokado}$^{1,2}$ 
\footnote{Electronic mail: tskokad@ipc.shizuoka.ac.jp}
and 
Kikuo~\textsc{Harigaya}$^{2,3}$
\footnote{Electronic mail: k.harigaya@aist.go.jp}
}
\address{
$^1$Faculty of Engineering, Shizuoka University, 
Hamamatsu 432-8561, Japan\\
$^2$Nanotechnology Research Institute, AIST, Tsukuba 305-8568, Japan \\
$^3$Synthetic Nano-Function Materials Project, AIST, Tsukuba 305-8568, Japan
}
\date{\today}

\maketitle

\begin{abstract}
As a novel function of ferromagnet (FM)/spacer/FM junctions, 
we theoretically investigate 
multiple-valued (or multi-level) cell property, 
which is in principle 
realized by sensing conductances of four states 
recorded with magnetization configurations of two FMs; 
(up,up), (up,down), (down,up), (down,down). 
In order to sense all the states, 
4-valued conductances corresponding to 
the respective states are necessary. 
We previously proposed that 
4-valued conductances are obtained in 
FM1/spin-polarized spacer (SPS)/FM2 junctions, 
where FM1 and FM2 have different spin polarizations, 
and the spacer depends on spin 
[J. Phys.: Condens. Matter {\bf 15}, 8797 (2003)]. 
In this paper, 
an ideal SPS is considered as 
a single-wall armchair carbon nanotube encapsulating magnetic atoms, 
where the nanotube shows on-resonance or off-resonance at the Fermi level 
according to its length. 
The magnitude of the obtained 4-valued conductances has 
an opposite order between the on-resonant nanotube and the off-resonant one, 
and this property can be understood 
by considering electronic states of the nanotube. 
Also, the magnetoresistance ratio between (up,up) and (down,down) 
can be larger than the conventional one 
between parallel and anti-parallel configurations. 
\end{abstract}

\narrowtext

\newpage
\noindent
\section{Introduction}
Spin polarized junctions (SPJ) 
such as ferromagnet(FM)/spacer/FM junctions~\cite{TMR,60,semi,delta} 
have been recently applied to 
elements in magnetic random access memories (MRAM) 
because of their magnetoresistance (MR) effect, 
which appears when an applied magnetic field changes an angle 
between magnetizations of two FMs. 
In the practical use, a spin-valve type, 
which corresponds to a memory cell of 2 values (1bit) 
represented by ``0'' and ``1''~\cite{spin-valve}, is usually adopted. 
For the writing process, 
the magnetization of only one side of the FM is changed 
under the applied field 
so that magnetization configurations 
between two FMs are parallel (P) or anti-parallel (AP). 
For the reading process, 
we use the difference in resistance between the P and AP configurations, 
the so-called MR effect.

For the SPJ, 
much effort has been made to develop 
elements in high density memories, 
which also have high sensitivity for the reading process. 
Important factors are considered to be 
(a) large MR ratio, 
whose expression in the unit of percent is defined by 
$100 \times (\Gamma_P - \Gamma_{AP})/\Gamma_{AP}$ with $\Gamma_{P(AP)}$ being 
conductance of the P (AP) case, 
and (b) ultra small junctions. 
The realistic SPJ for (a) can be, for example, 
Co-Fe/Al-O/Co-Fe junctions with the MR ratio of 60\% 
at room temperature~\cite{60}, 
epitaxially grown Ga$_{1-x}$Mn$_x$As/AlAs/Ga$_{1-x}$Mn$_x$As junctions 
with the MR ratio more than 70\% at 8 K~\cite{semi}, 
and Co/Fe-doped Al$_2$O$_3$/Ni$_{80}$Fe$_{20}$ junctions, 
whose MR ratio was enhanced as compared to 
the case of an undoped Al$_2$O$_3$~\cite{delta}. 
On the other hand, candidates of the SPJ for (b) may be 
FM/nanowire/FM junctions and 
FM/carbon nanotube/FM junctions~\cite{Tsuka,Zhao,Zhao2,Kim,Mehrez,Kokado}, 
where it should be noted that the nanotube length is 
about 200 nm~\cite{Zhao,Zhao2} or 250nm~\cite{Tsuka} 
in actual experiments 
and their size is very large at the present moment. 
In the further, however, we appear to see many problems with 
how to miniaturize the junctions, 
because challenges for the limits of the junctions size 
will be more and more severe 
in spite of the progress of experimental techniques.

Studies from another viewpoints far from the miniaturization 
should be necessary simultaneously. 
In fact, for other memories~\cite{set1} such as the Flash memory~\cite{MLC}, 
multiple valued (or multi-level) cell property~\cite{MLC}, 
which allows the plural bits to be stored in each memory cell 
and accordingly 
reduces the memory cell size by 1/(the number of bits), 
has been extensively studied. 
On the other hand, there are few studies for the SPJ. 
If such a property is included in the SPJ, 
they will function as an memory cell, 
which is more efficient than the conventional one.

We here describe 
the multiple-valued cell property in the general SPJ, 
based on our previous study~\cite{Kokado1}. 
In a possible scheme, 
recorded states are supposed to be 
four magnetization configurations of two FMs consisting of 
(up,up), (up,down), (down,up), (down,down), 
which are obtained by applying the magnetic fields to the respective FMs. 
The junctions correspond just to 2bits memory cells. 
Then, in order to sense all the states, 
4-valued conductances corresponding to the respective states are 
obviously necessary. 
By paying attention to the magnitude of total magnetization 
in the whole system, 
a model to obtain such conductances is considered to be 
FM1/spin-polarized spacer (SPS)/FM2 junctions, 
where the FM1 and FM2 have different spin polarizations~\cite{sp_ex}, 
according to the following procedure.  
First, a difference of conductances 
between (up,up) and (down,down) will appear, 
if the spacer is the SPS, 
where magnetization in the spacer is pinned. 
Second, a difference between (up,down) and (down,up) 
will be obtained by introducing the FM1 and FM2, 
in addition to the above mentioned SPS.

In this paper, 
as an ideal SPS to observe 4-valued conductances, 
we adopt a single-wall armchair carbon nanotube 
encapsulating magnetic atoms~\cite{nt1,nt2}, 
and investigate the spin dependent transport of 
``FM1/single-wall armchair carbon nanotube 
encapsulating magnetic atoms/FM2'' junctions 
using the Green's function technique. 
When all spins of magnetic atoms are pinned 
parallel to the magnetization axes of FM1 and FM2, 
4-valued conductances are explicitly obtained 
at a certain value of an exchange interaction between 
conduction electron spin and spin of magnetic atoms. 
The magnitude of the obtained 4-valued conductances has an opposite order 
between the nanotube with on-resonant behavior 
and that with off-resonant behavior. 
Also, the MR ratio between $\Uparrow,\Uparrow$ and $\Downarrow,\Downarrow$ 
can be larger than the conventional one between P and AP configurations.

\section{ideal spin polarized spacers to obtain 4-valued conductances}
We first consider ideal SPSs 
to certainly observe 4-valued conductances. 
In the previous study~\cite{Kokado1}, 
we found that 
4-valued conductances tend to be obtained 
in the case of largely spin-polarized spacer with 
\begin{eqnarray}
\frac{T_{\uparrow,\uparrow}-T_{\downarrow,\downarrow}}{T_{\uparrow,\uparrow}} 
\sim 1, 
\end{eqnarray}
where $T_{\uparrow,\uparrow}$ and $T_{\downarrow,\downarrow}$ are 
transmission coefficients of spin-up and spin-down, respectively, 
in an expression of the conductance~\cite{theory,kokado}, 
\begin{eqnarray}
\label{cond}
&&\Gamma = 
\frac{4 \pi^2 e^2 }{h} \sum_{\sigma=\uparrow,\downarrow}
\sum_{\sigma'=\uparrow,\downarrow}
T_{\sigma,\sigma'}
D_{1,\sigma}(E_{\mbox{\tiny F}}) 
D_{2,\sigma'}(E_{\mbox{\tiny F}}), 
\end{eqnarray}
with 
$\sigma$ (=$\uparrow$ or $\downarrow$) being spin of the conduction electron, 
$D_{1(2),\sigma}(E)$ being the local density-of-states (DOS) 
at an interfacial layer in FM1(2) 
at the Fermi level $E_{\mbox{\tiny F}}$, 
and $T_{\sigma,\sigma'}$ 
being a spin dependent transmission coefficient 
including spin-flip process of $\sigma \ne \sigma'$. 
Based on the fact, 
we give four objectives for the spacer having magnetic atoms, which are 
\begin{itemize}
\item[(i)] to strongly pin the magnetization 
of magnetic atoms in the spacer 
parallel to magnetization axes of FMs, 
\item[(ii)] to diminish magnetic couplings 
between magnetic atoms and FMs, 
\item[(iii)] to make effective couplings 
between the conduction electron spin and spins of magnetic atoms, 
which act as much as possible, 
\item[(iv)] to have a long spin-flip scattering length 
to conserve the spin of conduction electron through the spacer. 
\end{itemize}
As for (i), we propose that 
a coercive field of magnetic particles consisting of magnetic atoms 
is much higher than ones of FMs. 
For (ii), 
we suppose that 
distances between magnetic atoms and FMs 
should be controlled 
so that 
magnetic dipole-dipole interactions between them 
become very small and 
give little influence on the spin dependent conduction. 
In (iii), we should make a situation so that 
electrons in all conduction routes can 
interact with spins of magnetic atoms. 
For (iv), it is desired that 
the spin-orbit interaction of a material of the spacer 
is very small.

As a realistic SPS, which could satisfy such objectives, 
we bear in mind of 
a carbon nanotube encapsulating magnetic atoms~\cite{nt1,nt2} 
by the following reasons: 
First, the size of a particle consisting of magnetic atoms 
may be close to 
that of a single domain particle 
by tuning conditions of fabrication, 
where the single domain particle has a high coercive field. 
The coercive field was recently observed as about 0.5 kOe at 300 K 
even for Fe particles encapsulated 
with non-single domain size of about 70 nm~\cite{nt1}. 
Furthermore, for only the magnetic particle, 
it was experimentally shown that 
the Fe$_2$O$_3$ nanoparticle with the diameter of 6.3 nm 
has the coercive field of 1 kOe at 300 K~\cite{kimi}. 
If such the particle can be encapsulated in the nanotube, 
the nanotube will act as the ideal SPS. 
Second, magnetic atoms can be encapsulated 
in the inner region of the nanotube~\cite{nt1}, 
where 
distances between the atoms and nanotube edges 
could be tuned 
by controlling nanotube growth processes. 
Third, 
since the encapsulated magnetic atoms are completely surrounded 
by carbon atoms, 
electrons in all conduction routes can 
interact with spins of magnetic atoms. 
Here, magnitude of couplings may be also tunable 
by controlling nanotube growth processes, 
because they strongly depend on distances 
between carbon atoms and magnetic atoms. 
Fourth, 
the nanotube itself has very long spin-flip scattering lengths 
which extend to 130 nm at least~\cite{Tsuka}.

\section{``FM1/Carbon Nanotube 
Encapsulating Magnetic Atoms/FM2'' Junctions}
\subsection{Model and method}
We focus on ``FM1/single-walled armchair carbon nanotube 
encapsulating magnetic atoms/FM2'' junctions, 
where magnetic atoms are located in the center of the nanotube. 
Figure 1 shows a simplified model, 
in which the FM has a simple cubic structure, 
the $x$-direction of FMs is set to be semi-infinite, 
and their $yz$-directions have the periodic boundary condition 
by being regarded as an infinite system. 
The armchair nanotube has a finite length, and then 
it is regarded just as an armchair ribbon~\cite{Waka} 
with short periodicity. 
The each edge carbon atom of the nanotube is assumed to
interact with its nearest atom of the cubic lattice of the FM. 
On the other hand, the encapsulated magnetic atoms have localized spins. 
We here assume that 
their spins are divided into several spin clusters 
which interact with their nearest carbon atoms respectively, 
and all the spin clusters have the identical total spin. 
The total spin is approximately represented by 
a classical spin ${\bf S}$ 
on the assumption that the spin cluster consists of many spins, 
and further every ${\bf S}$ is set to have the same direction. 
Now, each ${\bf S}$ is arrayed along two dimer lines of the nanotube, and 
it has one to one interaction with carbon atoms in their lines. 
According to the theory on magnetic impurity problem~\cite{Schri}, 
we take into account antiferromagnetic exchange interactions 
between conduction electron spins and ${\bf S}$'s, 
where all the exchange interactions are here set to be same. 
Also, interactions between 
${\bf S}$'s and spins of FMs are neglected 
by assuming that 
they have little influence on the spin dependent conduction, 
under a balance 
between their coercive fields and magnetic fields 
due to respective spins. 
The balance is actually sensitive to 
distances between magnetic atoms and FMs.

Using a single orbital tight-binding model 
with nearest neighbor transfer integrals, 
the Hamiltonian is given by,
\begin{eqnarray}
\label{ham} 
&&H_{total}=H + V, \\
&&H =H_1^0 + H_2^0 + H_{NT}, \\
&& H_{NT} = H_{NT}^0 + H_{mag}, 
\end{eqnarray}
with
\begin{eqnarray}
\label{ham1} 
&& H_u^0=
   \sum_{i \in u} \sum_\sigma e_{i,\sigma} c_{i,\sigma}^\dag c_{i,\sigma} 
   +\sum_{\langle i,j \rangle \in u} \sum_\sigma 
     \left( t_{i,j} c_{i,\sigma}^\dag c_{j,\sigma} +{\rm h.c.} \right), \nonumber \\ 
\end{eqnarray}
for $u$=1, 2, $NT$, 
\begin {eqnarray}
&&H_{mag}= 
\sum_{i \in mag} \left( 
\Delta e \sum_\sigma c_{i,\sigma}^\dag c_{i,\sigma} 
-J \sum_{\sigma,\sigma'} 
{\bf \sigma}_{\sigma,\sigma'} \cdot {\bf S} 
c_{i,\sigma}^\dag c_{i,\sigma'} \right), \nonumber \\
\\
&&V=\sum_{\langle i,j \rangle} \sum_\sigma 
     \left( v_{i,j} c_{i,\sigma}^\dag c_{j,\sigma} +{\rm h.c.} \right),  
\end{eqnarray}
where 
$H_{1(2)}^0$ is Hamiltonian for the FM1 (FM2), 
$H_{NT}^0$ is that for the carbon nanotube, 
and $H_{mag}$ denotes interactions 
between the encapsulated magnetic atoms and 
the respective nearest carbon atoms, where 
$\sum_{i \in mag}$ means that the summation is taken for 
carbon atoms interacting with magnetic atoms. 
The term $V$ represents couplings between FMs and the nanotube, 
where 
the each edge carbon atom of the nanotube couples to 
its nearest atom of the cubic lattice of the FM. 
Here, $c_{i,\sigma}$ ($c_{i,\sigma}^{\dag} $) is 
the annihilation (creation) operator of an electron 
with spin-$\sigma$ ($=\uparrow {\rm or} \downarrow$) at the $i$-th site, 
$t_{i,j}$ and $v_{i,j}$ are 
transfer integrals between the $i$-th site and the $j$-th site, 
and $e_{i,\sigma}$ denotes the on-site energy 
for spin-$\sigma$ at the $i$-th site. 
Furthermore, 
$\Delta e$ is the difference of energy between 
the pure carbon atom and the carbon atom interacting with 
the magnetic atoms, 
$J$ is the antiferromagnetic exchange integral 
with negative sign~\cite{Schri}, 
and ${\bf \sigma}_{\sigma,\sigma'}$ is 
the $(\sigma,\sigma')$ component of the Pauli matrix 
for the conduction electron spin. 
Also, 
${\bf S}$ $[=(S_{x},S_{y},S_{z})]$ represents 
the classical spin with $S \equiv |{\bf S}|$.

Within the Green's function technique~\cite{theory,kokado}, 
we calculate the conductance at zero temperature, which is written by, 
\begin{eqnarray}
&&\Gamma =\frac{4 \pi^2 e^2 }{h}
{\rm Tr}[\hat{D}_1 \hat{T}^\dag \hat{D}_2 \hat{T}], \\
&&\hat{T}=V + VG^\dag V, \\
&&G=(E_{\mbox{\tiny F}} + {\rm i} 0 - H_{total})^{-1},
\end{eqnarray}
with
\begin{eqnarray}
&&\hat{D}_u=-\frac{1}{\pi} {\rm Im} G_u^0, \\
&&G_u^0=(E_{\mbox{\tiny F}} + {\rm i} 0 - H_u^0)^{-1}, 
\end{eqnarray}
for $u$=1, 2. 
Here, $\hat{D}_{1(2)}$ 
is the density-of-states (DOS) operator at $E_{\mbox{\tiny F}}$ 
of the FM1(2), and 
$\hat{T}$ is the $T$-matrix. 
We below name $\Gamma$ for the respective magnetization configurations as 
$\Gamma_{m1,m2}$, where $m1$ ($m2$) is the magnetization state of FM1 (FM2), 
which is $\Uparrow$ or $\Downarrow$.

Using these conductances, 
we obtain the MR ratio in the unit of percent, which is defined by 
\begin{eqnarray}
\label{MR}
R_{m1,m2} = 100 \times 
\frac{\Gamma_{\Downarrow,\Downarrow}-\Gamma_{m1,m2}}
{\Gamma_{\Downarrow,\Downarrow}}, 
\end{eqnarray}
where $m1$ ($m2$) is the magnetization state of FM1 (FM2) 
with $\Uparrow$ or $\Downarrow$.

In this calculation, we choose parameters as follows: 
We set $t_{i,j}=t~(<0)$~\cite{transf} and $v_{i,j}=0.1t$, 
assuming that $v$ is smaller than $t$ 
because of different types between two orbitals, 
imperfect lattice matches at the interface, and so on. 
When $E_{\mbox{\tiny F}}$=0, 
$e_{i,\uparrow}/|t|$ ($e_{i,\downarrow}/|t|$) 
is 5.1 (5.7) for the FM1 and 5.175 (5.625) for the FM2~\cite{comment}, 
by considering that the $s$-orbital, 
which is spin-polarized by coupling to the localized $d$-orbitals, 
contributes to the transport of the FMs. 
The spin polarization~\cite{sp_ex} at the interfacial layer 
of the FM1 (FM2) at $E_{\mbox{\tiny F}}$, $P_1$ ($P_2$), 
is then evaluated to be about 0.45 (0.25)~\cite{sp}. 
For the carbon atom, 
$e_{i,\uparrow}/|t|$ ($e_{i,\downarrow}/|t|$) is set to be 0 (0) 
by focusing on its $\pi$ orbital. 
By taking into account that the $\pi$ orbital is 
coupled to $d$-orbitals of the magnetic atoms, whose energies are lower 
than $e_{i,\uparrow}/|t|$ 
and $e_{i,\downarrow}/|t|$ of the $\pi$ orbital, 
$\Delta e/ |t|$ is considered to be positive and it is put as 0.2. 
Also, 
the number of unit cells in the circumference direction of the nanotube 
is 10. 
The number of dimer lines~\cite{Waka} $N$ is 20, 21, and 22, 
which correspond to nanotubes with 
the on-resonance, the off-resonance, and the off-resonance 
at $E_{\mbox{\tiny F}}$, respectively~\cite{Mehrez,Waka}. 
It is well known that 
the armchair nanotube shows the on-resonant behavior for $N=3M-1$, 
the off-resonant one for $N=3M$, and the off-resonant one for $N=3M+1$, 
respectively, with $M$ being an integer~\cite{Mehrez,Waka}. 
For $N$=20 and 21, 
the spin, ${\bf S}$, is arrayed at each carbon atom 
in 10-th and 11-th dimer lines of the nanotube, 
while it is arrayed at each carbon atom 
in its 11-th and 12-th dimer lines for $N$=22. 
In each $N$, the number of ${\bf S}$'s is 40, which is obtained from 
the relation of 
(the number of unit cells in the circumference direction of the nanotube) 
$\times$ 
(the number of dimer lines with ${\bf S}$'s) 
$\times$ 
(the number of sublattices in the unit cell)=
10 $\times$ 2 $\times$ 2. 
Furthermore, ${\bf S}$'s are considered to exist parallel to $yz$-plane, 
and an angle between ${\bf S}$'s and $z$-axis is written as $\theta$.

\subsection{Calculated results and considerations}
In the upper panel of Fig. 2(a), we show 
$-JS/|t|$ dependence of the conductance, 
$\Gamma_{m1,m2}$, for $\theta$=0 in the case of $N$=20. 
At $JS/|t|$=0, 
$\Gamma_{m1,m2}$ has a difference 
only between P and AP configurations. 
For $JS/|t|\ne$0, 
differences among all conductances 
appear, 
and become largest in the vicinity of $JS/|t|$=$-$0.2. 
Also, $\Gamma_{m1,m2}$ has a relation of 
$\Gamma_{\Downarrow,\Downarrow}> 
\Gamma_{\Uparrow,\Downarrow}>
\Gamma_{\Downarrow,\Uparrow}>
\Gamma_{\Uparrow,\Uparrow}$ 
for a wide range of $-0.8<JS/|t|< 0$.

The middle and lower panels of Fig. 2(a) show 
$\Gamma_{m1,m2}$ for $\theta$=0 
in the case of $N$=21 and 22, respectively. 
In the case of $N$=21, 
differences among all conductances 
become large in the vicinity of $JS/|t|$=$-$0.4, 
while they do in the vicinity of $JS/|t|$=$-$0.75 in the case of $N$=22. 
It should be noted that 
$\Gamma_{m1,m2}$ of $N$=21 and 22 has 
a relation of $\Gamma_{\Uparrow,\Uparrow}> 
\Gamma_{\Downarrow,\Uparrow}>
\Gamma_{\Uparrow,\Downarrow}>
\Gamma_{\Downarrow,\Downarrow}$ for a wide range of $-0.8< JS/|t| < 0$, 
and its order is opposite to that of $N$=20.

The behavior of conductances can be understood by considering 
electronic states in the center of the nanotube, 
where magnetic atoms are encapsulated. 
We therefore investigate the local DOS, defined by, 
\begin{eqnarray}
{\rm Local~DOS}=-\frac{1}{\pi} {\rm Im} \sum_{i \in mag} \langle i| 
(E_{\mbox{\tiny F}} + {\rm i} 0 - H_{NT})^{-1} | i \rangle,
\end{eqnarray}
where 
$|i\rangle$ represents the orbitals of 
carbon atoms interacting with magnetic atoms, and 
$\sum_{i \in mag}$ means that the summation is taken for 
those carbon atoms.

In the upper, middle, lower panels of Fig. 3(a), we show the local DOS 
for $\Delta e/|t|$=0 and $JS/|t|$=0, 
$\Delta e/|t|$=0.2 and $JS/|t|$=0, 
and $\Delta e/|t|$=0.2 and $JS/|t|$=$-$0.2, 
respectively, in the case of $N$=20. 
For $\Delta e/|t|$=0 and $JS/|t|$=0, i.e., the case of no encapsulated atoms, 
small peaks of spin-up and spin-down are found at $E_{\mbox{\tiny F}}$ (=0), 
as shown in the upper panel of Fig. 3(a). 
The feature just represents the on-resonance at $E_{\mbox{\tiny F}}$. 
The peaks originate from wave functions at the Gamma point~\cite{Waka}, 
which are delocalized over the whole nanotube. 
For $\Delta e/|t|$=0.2 and $JS/|t|$=0, i.e., 
the case of encapsulated nonmagnetic atoms, 
the peaks of spin-up and spin-down 
are shifted to higher energy with the same magnitude 
[see middle panel of Fig. 3(a)]. 
Then, 2-valued conductances are obtained 
because of no difference of local DOSs between spin-up and spin-down. 
On the other hand, 
for $\Delta e/|t|$=0.2 and $JS/|t|$=$-$0.2, i.e., 
the case of encapsulated magnetic atoms, 
the peak of spin-down is close to $E_{\mbox{\tiny F}}$ 
as shown in the lower panel of Fig. 3(a). 
Since the electron favors to transmit 
when the DOS is large around $E_{\mbox{\tiny F}}$, 
transmission of spin-down electron increases 
compared to that of spin-up electron. 
By taking into account the spin dependent DOSs of FMs 
with $P_1$=0.45 and $P_2$=0.25, 
$\Gamma_{\Downarrow,\Downarrow}$ ($\Gamma_{\Uparrow,\Uparrow}$) 
becomes largest (smallest) in all conductances, 
while $\Gamma_{\Uparrow,\Downarrow}$ and $\Gamma_{\Downarrow,\Uparrow}$ 
are present between 
$\Gamma_{\Downarrow,\Downarrow}$ and $\Gamma_{\Uparrow,\Uparrow}$, 
and 
$\Gamma_{\Uparrow,\Downarrow} > \Gamma_{\Downarrow,\Uparrow}$ 
is realized also.

The upper, middle, lower panels of Fig. 3(b) show the local DOS 
for $\Delta e/|t|$=0 and $JS/|t|$=0, 
$\Delta e/|t|$=0.2 and $JS/|t|$=0, 
$\Delta e/|t|$=0.2 and $JS/|t|$=$-$0.4, respectively, 
in the case of $N$=21. 
In the upper panel of Fig. 3(b), 
no peaks are found at $E_{\mbox{\tiny F}}$ 
for $\Delta e/|t|$=0 and $JS/|t|$=0. 
It represents the off-resonance at $E_{\mbox{\tiny F}}$. 
By setting $\Delta e/|t|$=0.2 and $JS/|t|$=0, 
peaks of spin-up and spin-down near $E_{\mbox{\tiny F}}$ 
are shifted to higher energy with the same magnitude 
as shown in middle panel of Fig. 3(b), 
and then 2-valued conductances are obtained. 
For $\Delta e/|t|$=0.2 and $JS/|t|$=$-$0.4, 
the peak of spin-up 
appears in the vicinity of $E_{\mbox{\tiny F}}$ 
[see the lower panel of Fig. 3(b)]. 
Therefore, based on the DOSs of the FMs, 
$\Gamma_{\Uparrow,\Uparrow}$ ($\Gamma_{\Downarrow,\Downarrow}$) 
is largest (smallest) in all conductances, 
while $\Gamma_{\Uparrow,\Downarrow}$ and $\Gamma_{\Downarrow,\Uparrow}$ 
are present between 
$\Gamma_{\Uparrow,\Uparrow}$ and $\Gamma_{\Downarrow,\Downarrow}$, 
and also $\Gamma_{\Downarrow,\Uparrow}>\Gamma_{\Uparrow,\Downarrow}$ 
is realized. 
The case of $N$=22 can be understood based on 
the local DOSs shown in Fig. 3(c), too.

We systematically understand the above peak shifts as follows: 
When the carbon atom interacts with the magnetic atoms, 
its energy levels of spin-up are shifted to high energy, 
while its energy levels of spin-down are not largely altered,  
owing to $\Delta e/|t| >0$ and $JS/|t| <0$. 
For the on-resonant nanotube, 
in which energy levels of spin-up and spin-down 
exist at $E_{\mbox{\tiny F}}$ for $\Delta e/|t|$=$JS/|t|$=0, 
the energy levels of spin-down 
are closer to $E_{\mbox{\tiny F}}$ than those of spin-up 
when $\Delta e/|t|>0$ and $JS/|t|<0$. 
On the other hand, for the off-resonant nanotube, 
in which no energy levels of spin-up and spin-down 
exist at $E_{\mbox{\tiny F}}$ for $\Delta e/|t|$=$JS/|t|$=0, 
the energy levels of spin-up near the top of the valence band 
approach to $E_{\mbox{\tiny F}}$ 
by setting $\Delta e/|t|>0$ and $JS/|t|<0$.

The MR ratio $R_{m1,m2}$ is shown in Fig. 2(b). 
In all panels, at $JS/|t|$=0, 
$R_{m1,m2}$ is finite only between P and AP configurations 
and $|R_{m1,m2}|$ is about 15\%. 
For a wide range of $-0.8 < JS/|t| < 0$, 
$|R_{m1,m2}|$ between P and AP configurations 
can be more than 15\%. 
It should be emphasized that 
$|R_{m1,m2}|$ between $\Uparrow,\Uparrow$ and $\Downarrow,\Downarrow$
is larger than the conventional one between P and AP configurations. 
This feature is consistent with results 
for the case of the largely spin-polarized spacer with 
$(T_{\uparrow,\uparrow}-T_{\downarrow,\downarrow})/
T_{\uparrow,\uparrow} \sim 1$ 
in the previous work~\cite{Kokado1}. 
Also, it is characteristic that 
for a wide range of $-0.8 < JS/|t| < 0$, 
$R_{m1,m2}$'s of $N$=20 and $N$=21 and 22 exhibit 
the positive MR and the negative one, respectively, 
reflecting the opposite order of $\Gamma_{m1,m2}$ between them.

In the following, we consider the $\theta$ dependence of 
$\Gamma_{m1,m2}$ and $R_{m1,m2}$, 
shown in Figs. 4(a) and (b), respectively. 
In each figure, the upper, middle, and lower panels are 
cases of $JS/|t|$=$-$0.2 of $N$=20, 
$JS/|t|$=$-$0.4 of $N$=21, 
and $JS/|t|$=$-$0.8 of $N$=22, respectively. 
A condition of $\theta/\pi$=$-$0.5 (0) represents 
that ${\bf S}$'s are oriented in the $-y$ ($z$) direction. 
In all panels, at $\theta/\pi=-0.5$, 
only the difference of $\Gamma_{m1,m2}$ 
between the P and AP configurations is present, 
and then $|R_{m1,m2}|$ 
between them takes small values of less than 5 \%, 
because of the spin-flip transmission. 
As $\theta$ approaches to 0, 
differences of $\Gamma_{m1,m2}$ 
among all configurations 
become large, 
because a difference of diagonal elements of $\hat{T}$ with regards to spin 
between spin-up and spin-down increases. 
Further, $|R_{m1,m2}|$ 
between $\Uparrow$,$\Uparrow$ and $\Downarrow$,$\Downarrow$ increases, too.

\section{Conclusion}
For 
``FM1/armchair carbon nanotube encapsulating magnetic atoms/FM2'' junctions, 
we theoretically investigated the multiple-valued cell property, 
which is in principle realized by sensing four states 
recorded with the magnetization configurations of two FMs. 
The obtained 4-valued conductances are strongly influenced by 
electronic states of the nanotube, 
and directions of spins of magnetic atoms. 
The order of magnitude of 4-valued conductances 
is opposite between the on-resonant nanotube and the off-resonant one. 
Furthermore, the MR ratio between (up,up) and (down,down) 
can be larger than the conventional one between P and AP configurations.

From the viewpoint of device applications, 
we expect that the present junctions will be 
a candidate for elements of the 2bits/cell MRAM. 
At the same time, 
the junctions with magnetization reversals 
between $\Uparrow,\Uparrow$ and $\Downarrow,\Downarrow$
may be more efficient magnetic sensor than 
the conventional spin-valve type~\cite{spin-valve}, 
in the case of the largely spin-polarized spacer.

\section{Acknowledgements}
This work has been supported by 
Special Coordination Funds for Promoting Science and Technology, Japan. 
One of the authors (K.H.) acknowledges 
the partial financial support from NEDO via 
Synthetic Nano-Function Materials Project, AIST, Japan, too.

\begin{figure}[ht]
\begin{center}
\epsfxsize=7cm \epsfbox{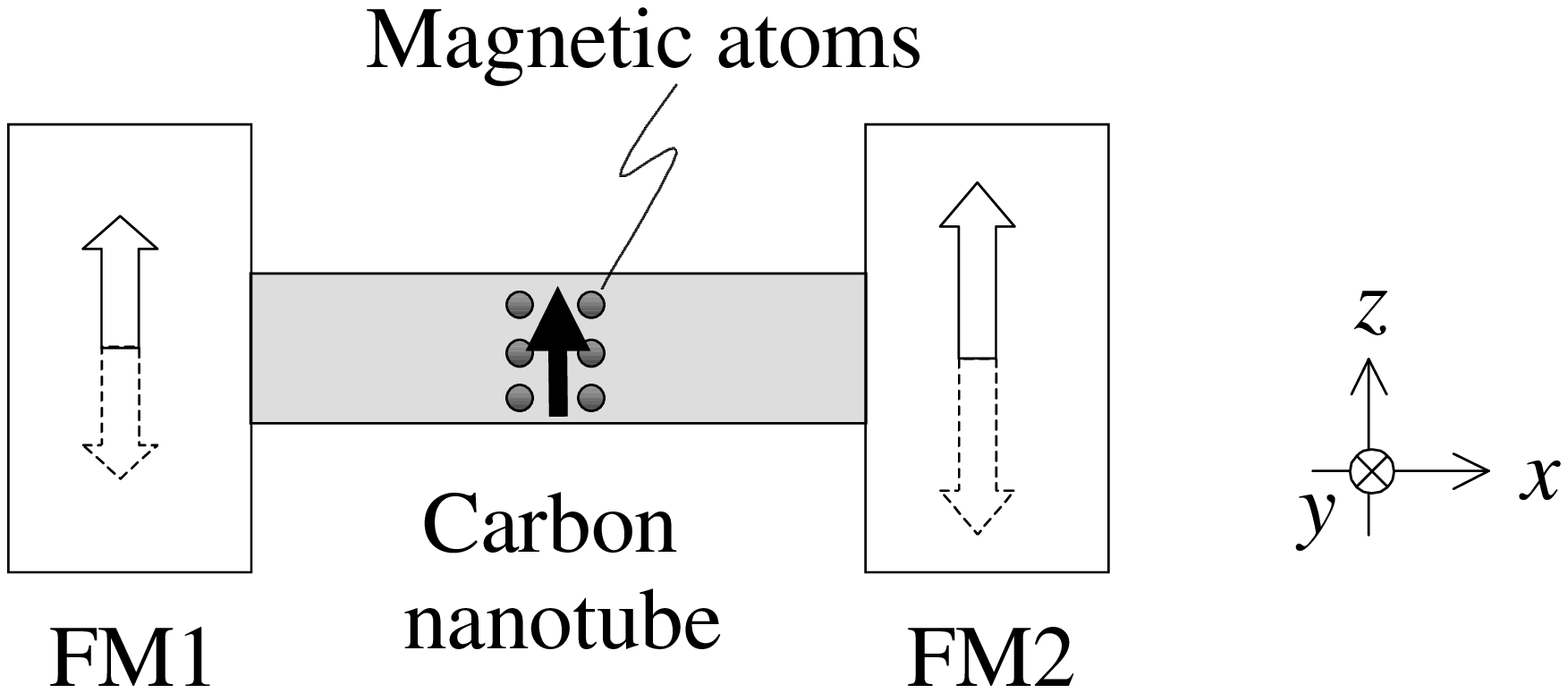}
\caption{
A schematic illustration 
of spin-polarized junctions with the carbon nanotube, 
where magnetic atoms are encapsulated in the center of the nanotube. 
Electric currents flow in the $x$-direction. 
}
\end{center}
\end{figure}

\newpage
\noindent
\begin{figure}[ht]
\begin{center}
\epsfxsize=6cm \epsfbox{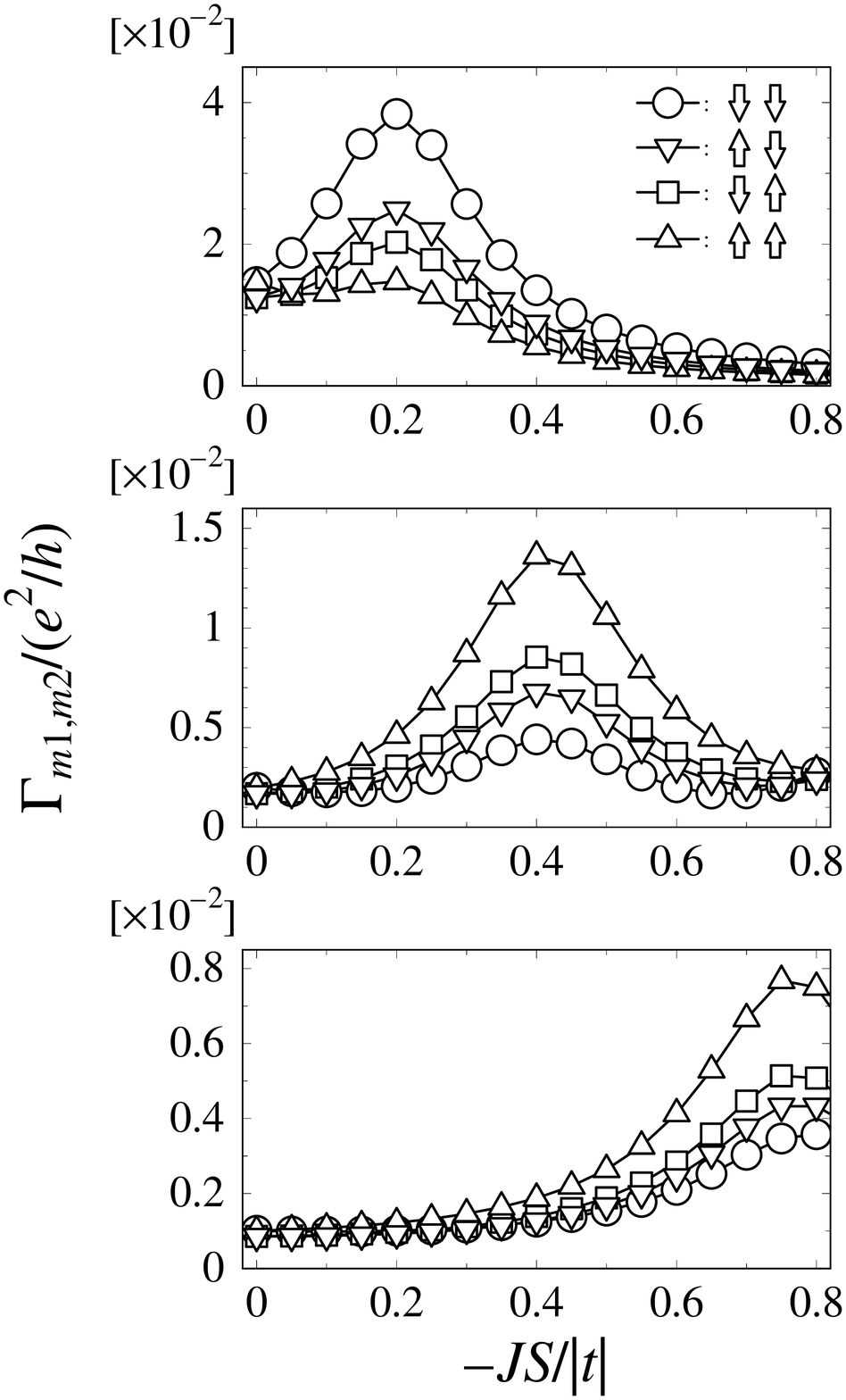}\\[-0.5cm]
\hspace*{-6cm}(a) \\
\vspace{0.5cm}
\epsfxsize=6cm \epsfbox{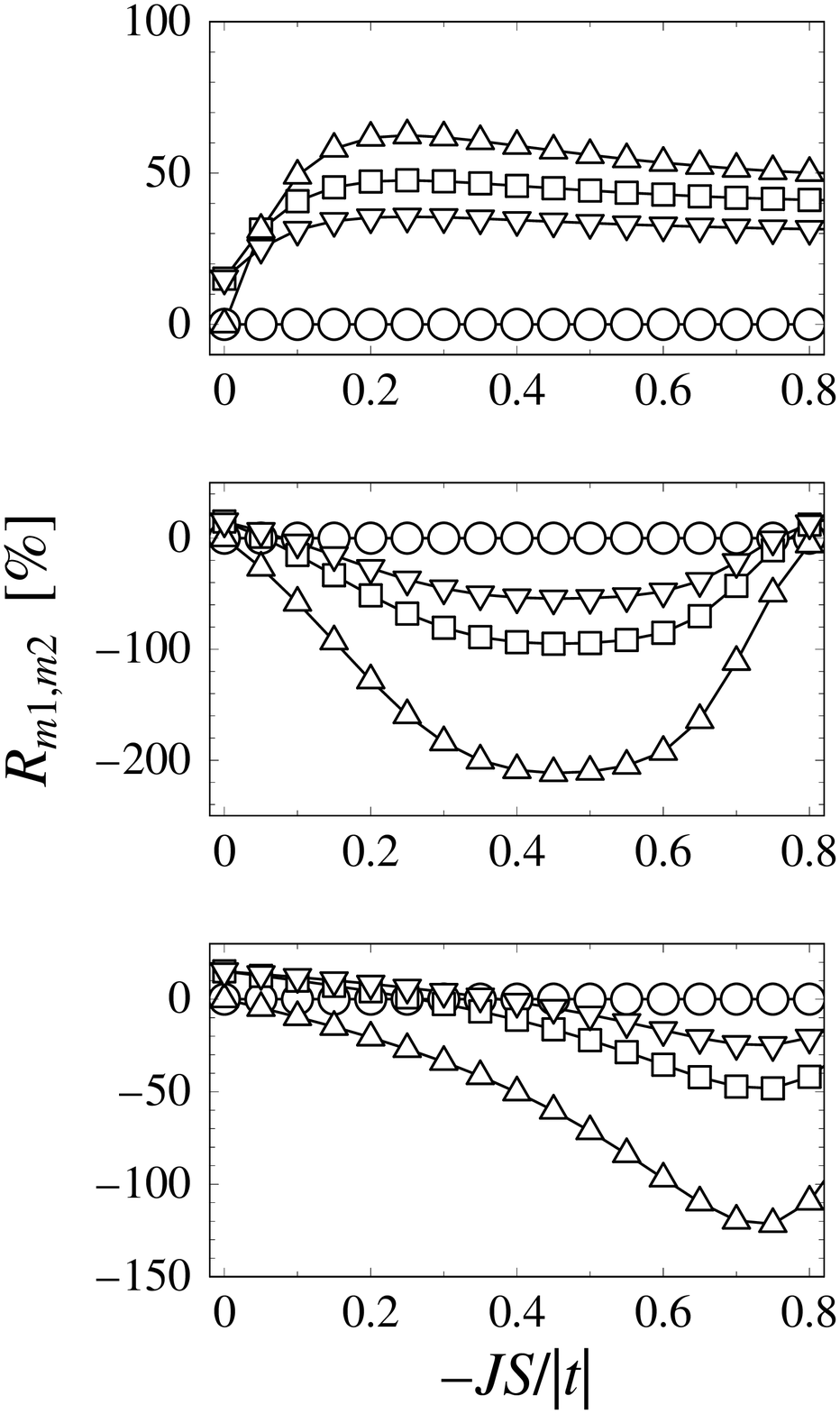}\\[-0.5cm]
\hspace*{-6cm}(b) \\
\caption{
(a) 
The conductance 
$\Gamma_{m1,m2}$ vs $-JS/|t|$ for $\theta$=0 and $\Delta e/|t|$=0.2. 
The meanings of dots in all panels are 
$\bigcirc$: $\Gamma_{\Downarrow,\Downarrow}$, 
$\bigtriangledown$: $\Gamma_{\Uparrow,\Downarrow}$, 
$\Box$: $\Gamma_{\Downarrow,\Uparrow}$, 
$\bigtriangleup$: $\Gamma_{\Uparrow,\Uparrow}$. 
(b) 
The MR ratio $R_{m1,m2}$ vs $-JS/|t|$ for $\theta$=0 and $\Delta e/|t|$=0.2. 
In each figure, the upper, middle, and lower panels are cases of 
$N$=20, 21, and 22, respectively. 
The meanings of dots in all panels are 
$\bigcirc$: $R_{\Downarrow,\Downarrow}$, 
$\bigtriangledown$: $R_{\Uparrow,\Downarrow}$, 
$\Box$: $R_{\Downarrow,\Uparrow}$, 
$\bigtriangleup$: $R_{\Uparrow,\Uparrow}$. 
}
\end{center}
\end{figure}

\newpage
\noindent
\begin{figure}[ht]
\begin{center}
\epsfxsize=4.5cm \epsfbox{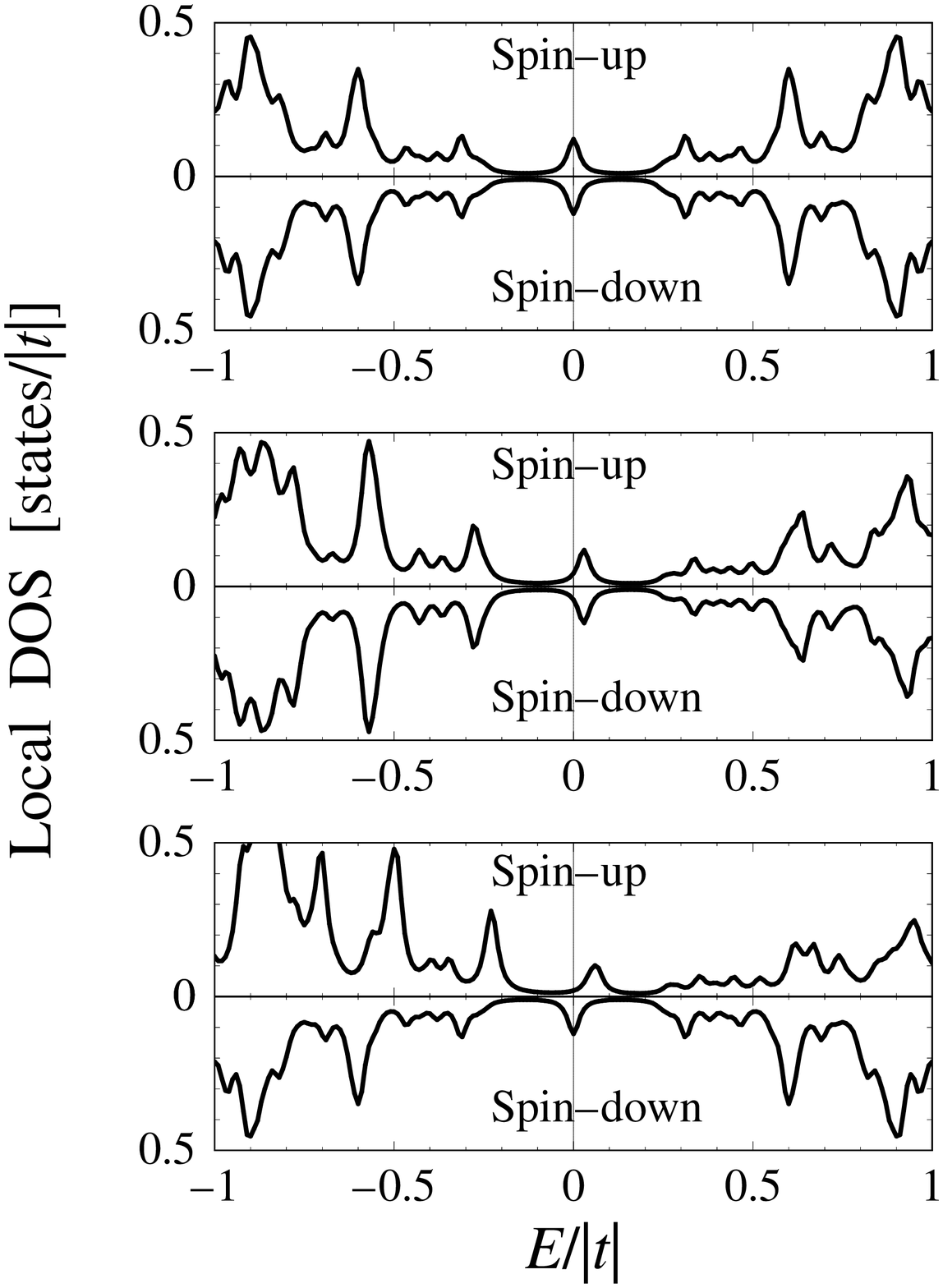}\\[-0.5cm]
\hspace*{-5cm}(a) \\
\vspace{0.5cm}
\epsfxsize=4.5cm \epsfbox{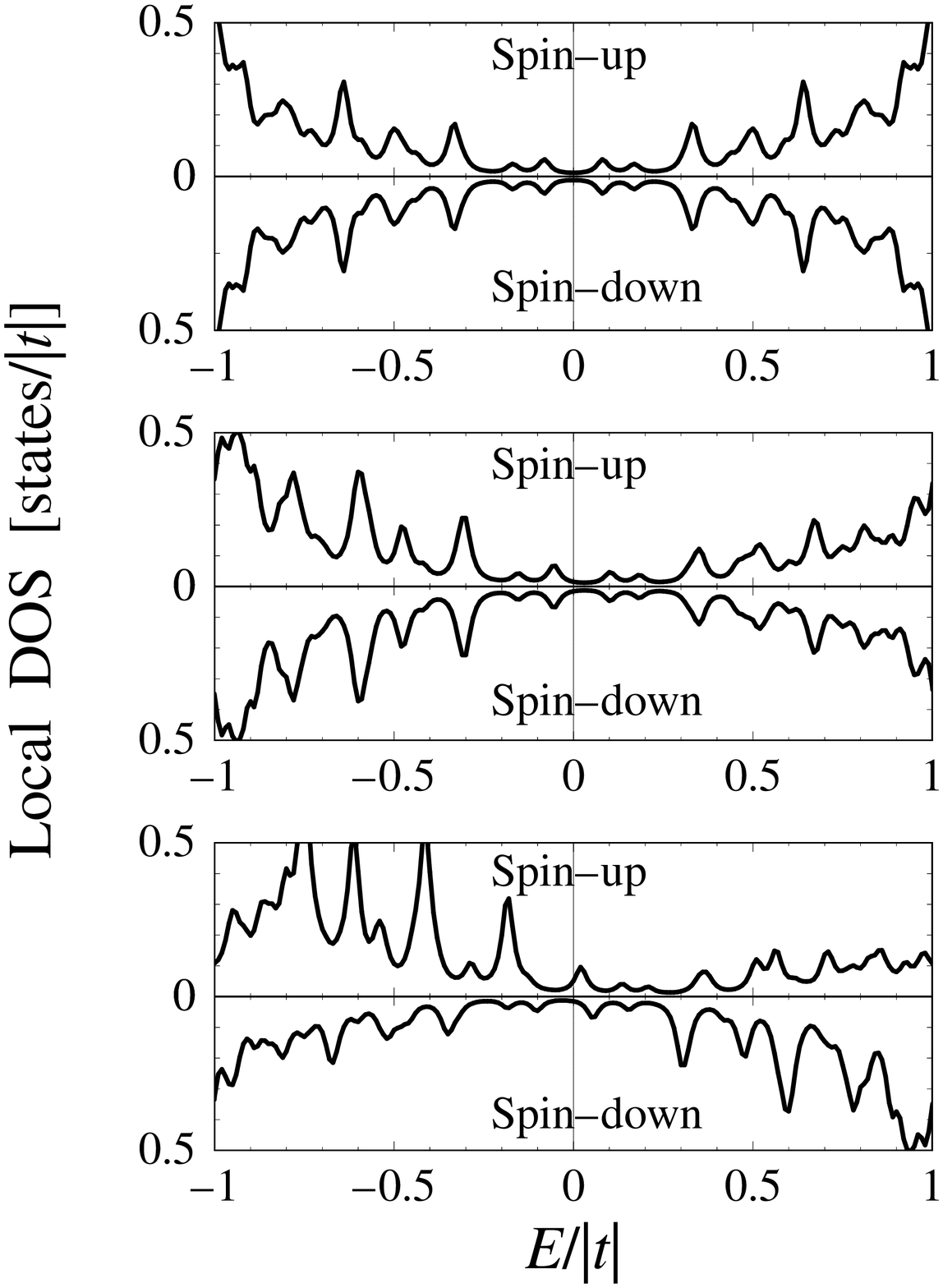}\\[-0.5cm]
\hspace*{-5cm}(b) \\
\vspace{0.5cm}
\epsfxsize=4.5cm \epsfbox{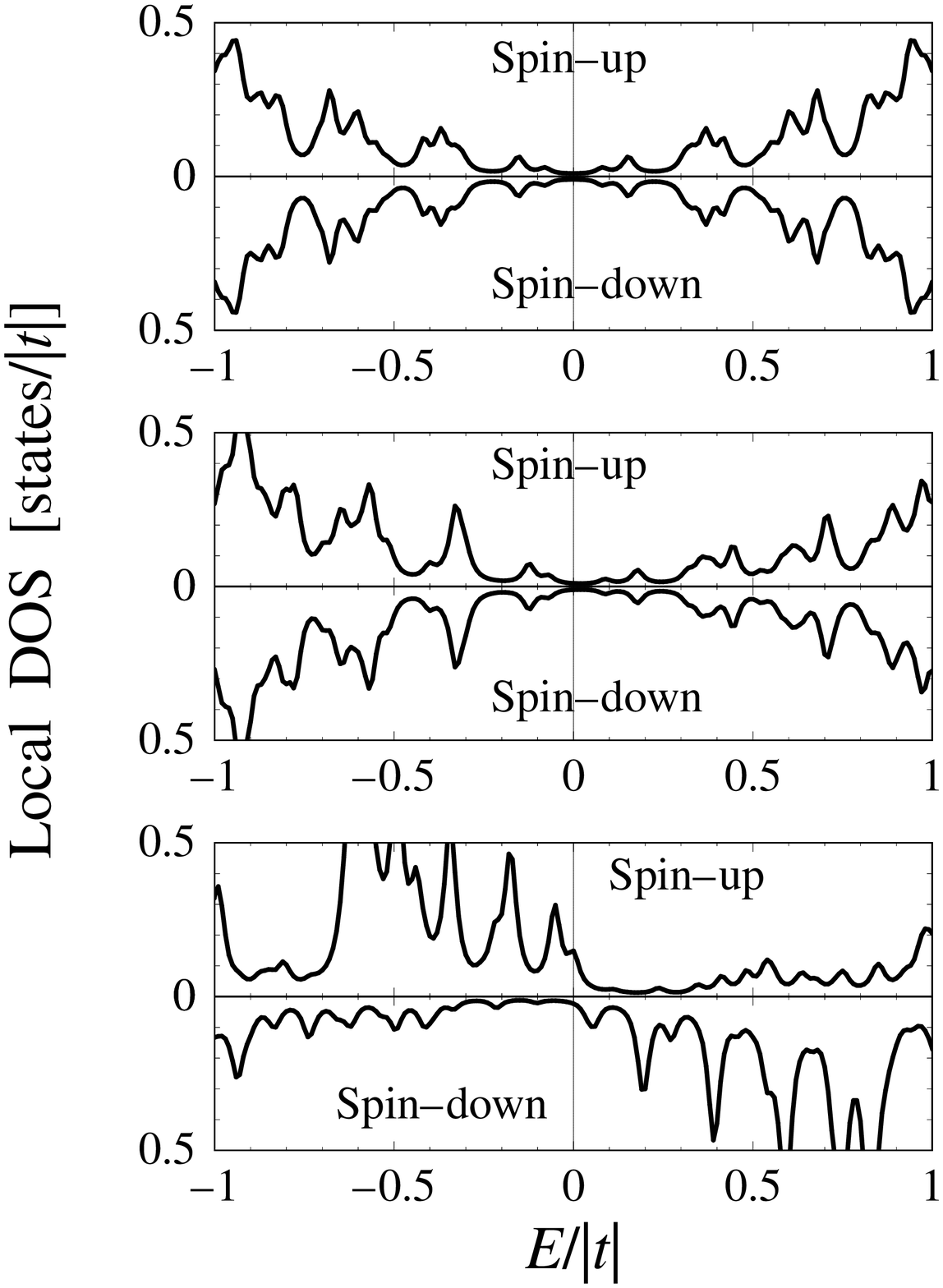}\\[-0.5cm]
\hspace*{-5cm}(c) \\
\vspace{0.5cm}
\caption{
The local DOS in the center of the nanotube, where 
magnetic atoms are encapsulated. 
(a) The case of $N$=20. 
Upper panel: $\Delta e/|t|$=0 and $JS/|t|$=0. 
Middle panel: $\Delta e/|t|$=0.2 and $JS/|t|$=0. 
Lower panel: $\Delta e/|t|$=0.2 and $JS/|t|$=$-$0.2. 
(b) The case of $N$=21. 
Upper panel: $\Delta e/|t|$=0 and $JS/|t|$=0. 
Middle panel: $\Delta e/|t|$=0.2 and $JS/|t|$=0. 
Lower panel: $\Delta e/|t|$=0.2 and $JS/|t|$=$-$0.4. 
(c) The case of $N$=22. 
Upper panel: $\Delta e/|t|$=0 and $JS/|t|$=0. 
Middle panel: $\Delta e/|t|$=0.2 and $JS/|t|$=0. 
Lower panel: $\Delta e/|t|$=0.2 and $JS/|t|$=$-$0.8. 
Here, $E_{\mbox{\tiny F}}$=0 is set.
}
\end{center}
\end{figure}

\newpage
\noindent
\begin{figure}[ht]
\begin{center}
\epsfxsize=6cm \epsfbox{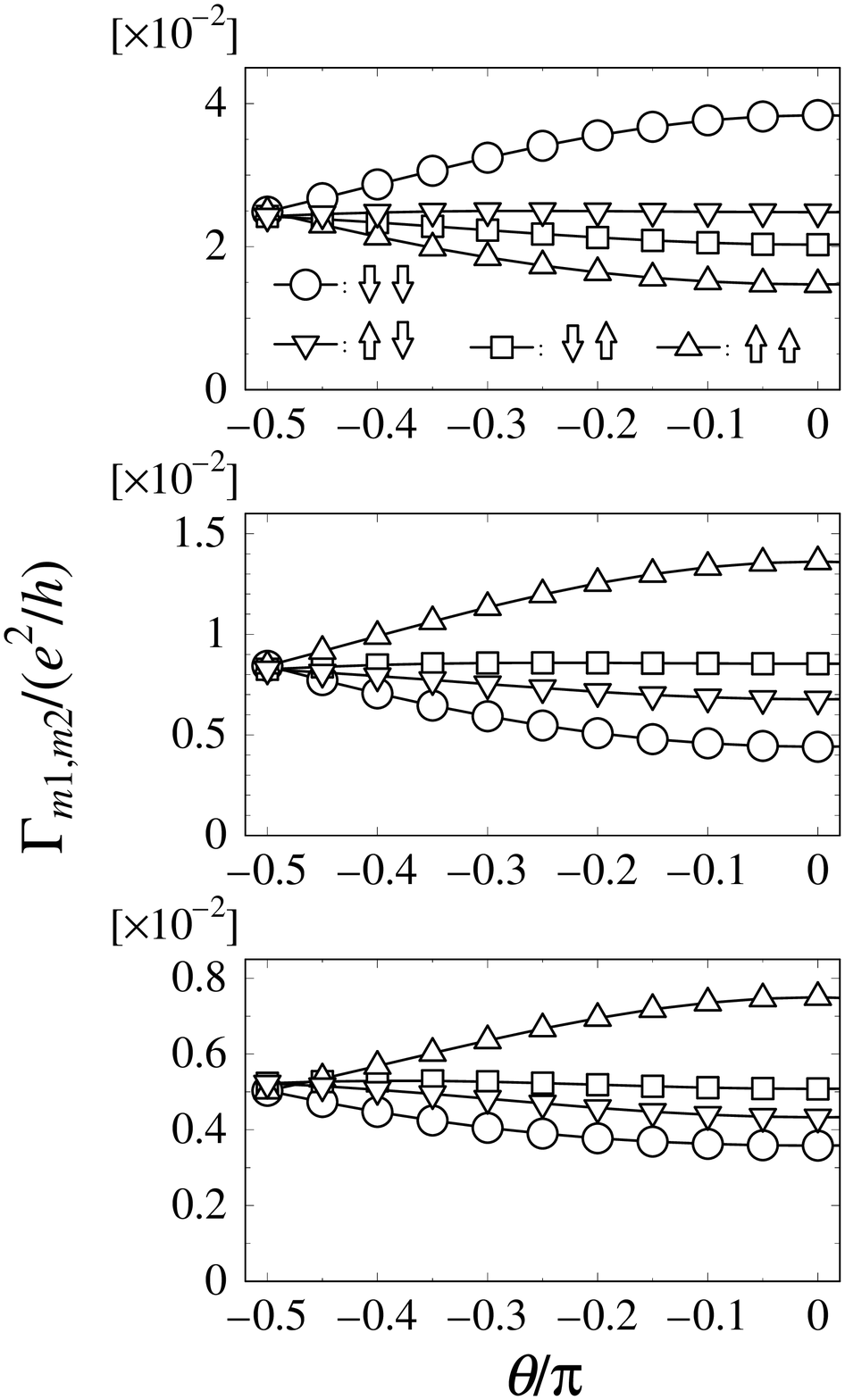}\\[-0.5cm]
\hspace*{-6cm}(a) \\
\vspace{0.5cm}
\epsfxsize=6cm \epsfbox{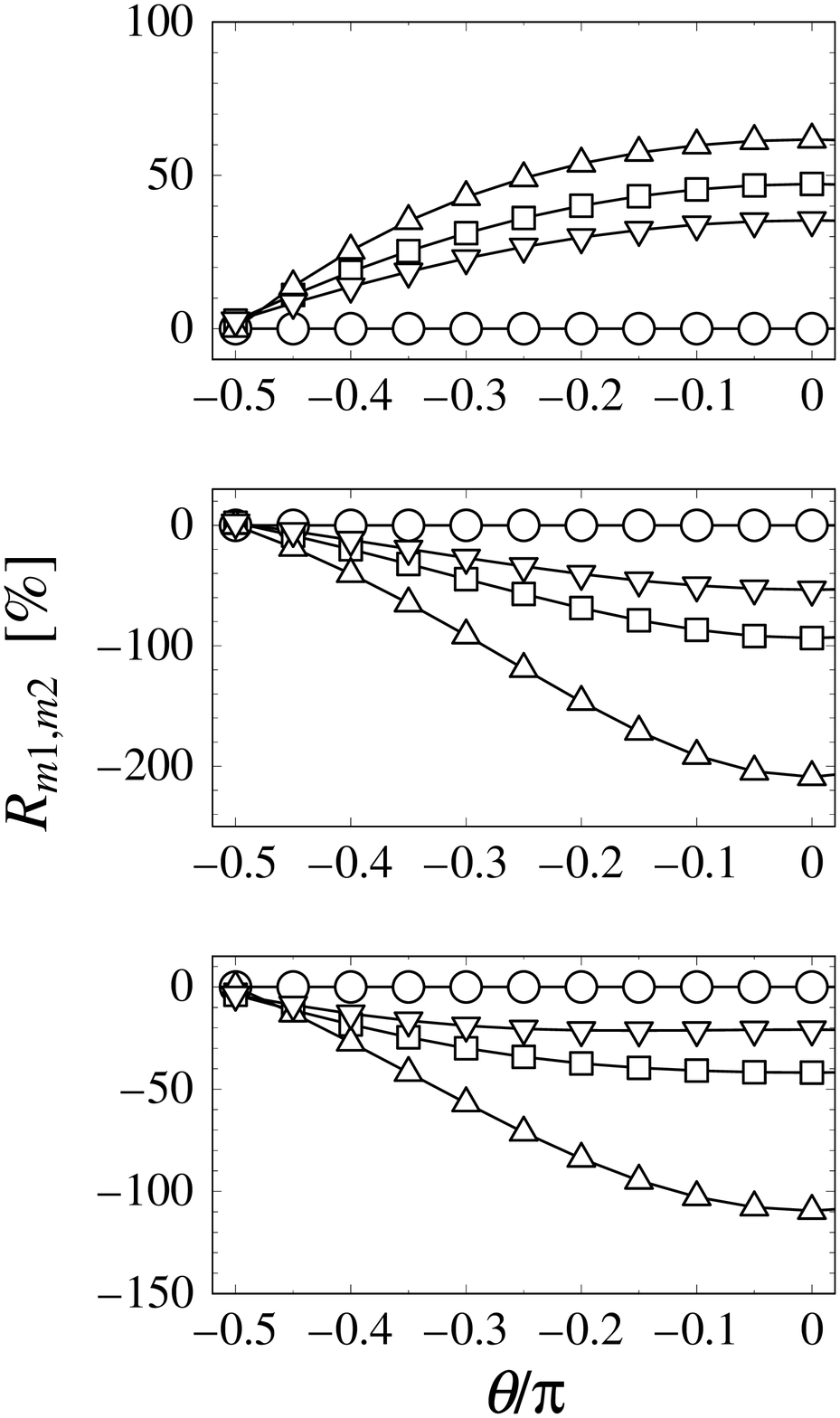}\\[-0.5cm]
\hspace*{-6cm}(b) \\
\caption{
(a) 
The conductance $\Gamma_{m1,m2}$ vs $\theta$. 
The meanings of dots in all panels obey those of Fig. 2(a). 
(b) 
The MR ratio $R_{m1,m2}$ vs $\theta$. 
In each figure, the upper, middle, and lower panels are 
cases of $JS/|t|$=$-$0.2 of $N$=20, 
$JS/|t|$=$-$0.4 of $N$=21, 
and $JS/|t|$=$-$0.8 of $N$=22, respectively. 
The meanings of dots in all panels obey those of Fig. 2(b). 
}
\end{center}
\end{figure} 


\begin{thebibliography}{9}
\bibitem{sp_ex}
The spin polarization is defined by $P_{1(2)}= 
(D_{1(2),\uparrow} - D_{1(2),\downarrow})/
(D_{1(2),\uparrow} + D_{1(2),\downarrow})$, 
where $D_{1(2),\uparrow}$ is the density-of-states (DOS) 
at the Fermi level of the FM1(2) for spin-up 
and $D_{1(2),\downarrow}$ is that of the FM1(2) for spin-down. 
Also, for the spin polarization at the interfacial layer, 
the DOS at the interfacial layer is used as $D_{1(2),\uparrow}$. 
For example, see 
J. M. De Teresa, A. Barthelemy, A. Fert, J. P. Contour, 
F. Montaigne, and P. Seneor, Science {\bf 286}, 507 (1999).
\bibitem{transf}
In the previous work~\cite{Mehrez}, $t$ was set to be $-$2.75 eV.
\end{thebibliography}
\end{document}